\documentclass[a4paper,12pt]{article}
\usepackage{epsfig,float}
\usepackage[left=2cm,right=2cm]{geometry}
\usepackage{a4wide,graphicx}

\newcommand{\eps}{\epsilon}

\begin{document}

\title{Scalar, axial-vector and tensor resonances from the $\rho D^*$, $\omega
D^*$
interaction in the hidden gauge formalism.}

\author{R. Molina$^1$, H. Nagahiro$^{2,\,3}$, A. Hosaka$^{3}$ and E. Oset$^1$}
\maketitle

\begin{center}
$^1$ Departamento de F\'{\i}sica Te\'orica and IFIC,
Centro Mixto Universidad de Valencia-CSIC,
Institutos de Investigaci\'on de Paterna, Aptdo. 22085, 46071 Valencia,
 Spain\\
$^2$ Department of Physics, Nara Women's University,
Nara 630-8506, Japan\\
$^3$ Research Center for Nuclear Physics (RCNP), Osaka University,
Ibaraki, Osaka 567-0047, Japan
\end{center}

\date{}

 \begin{abstract}  

We study composite systems of light ($\rho$ and $\omega$) and heavy
  ($D^*$) vector mesons by using the interaction within the hidden gauge
  formalism. 
We find a strong attraction 
in the isospin, spin channels $(I,S)=(1/2,0)$; $(1/2,1)$; $(1/2,2)$ with
  positive parity. The 
attraction is large enough to strongly bind these mesons in states with these
quantum numbers, leading to states which can be identified with
$D_2^*(2460)$ and probably with
$D^*(2640)$, the last one without 
 experimental spin and parity
assignment. In the case of $I=3/2$ one obtains
 repulsion and thus, no exotic mesons in this sector are generated in the
 approach.

\end{abstract}

\section{Introduction}

     Recently a study of the $\rho \rho$ interaction with the hidden gauge
formalism \cite{hidden1,hidden2,hidden3} was carried out in \cite{hidden3}.
The hidden gauge symmetry (HGS) was introduced by Bando-Kugo-Yamawaki
where the $\rho$ meson was regarded as a dynamical gauge boson of the
HGS of the non-linear sigma model.
One of interesting facts which is relevant in the present discussion is
that there is a strong attraction in the isospin and spin channels
$(I,S)=(0,0)$ and $(0,2)$, 
which
is
enough to bind the $\rho \rho$ system leading to a tensor and a scalar
meson which could be identified with the $f_0(1370)$ and $f_2(1270)$ meson
states \cite{raquel}. In a later work the radiative decay of these states in
$\gamma \gamma$, a channel with rates very sensitive to the nature of the
resonances, was studied \cite{junko} obtaining results in agreement with the 
PDG \cite{pdg} for
the case of the tensor state and in qualitative agreement with preliminary
results at Belle for the scalar state \cite{uehara}. The work of \cite{hidden3} has been recently extended to SU(3) for the interaction of the vectors of the nonet, where several states, which can be identified with existing resonances are also dynamically generated \cite{geng}. The success of the approach
encourages us to study the charm sector in  the lightest case, the
one for the interaction of the $\rho,\omega$ and $D^*$ mesons. Another possible approach to 
this work could be done following the lines of \cite{carmen} for meson-baryon interaction involving $SU(8)$ symmetry. Work in this direction is in progress \cite{Jgeng} but we can advance that while for pseudoscalar-baryon interaction the approaches are equivalent, when vector meson are involved the results are very different \cite{private}.

The starting point in our approachis the
interaction of vector mesons among themselves provided by the hidden gauge
formalism, which now has to be generalized to SU(4) to accommodate the charm
vectors $D^*$ into the framework. Admitting that SU(4) is more strongly broken
than SU(3), the SU(4) symmetry is invoked in the basic hidden gauge Lagrangians
but is already broken in the vector exchange diagrams that provide the
amplitudes for the vector-vector interaction. What we find is a strong
attraction in the isospin, spin channels
$(I,S)=(1/2,0);\,(1/2,1);\,(1/2,2);$ which leads to 
bound $\rho(\omega) D^*$ states in all these channels.  In the case of
$I=3/2$ we find 
repulsion and hence we do not generate states that would qualify as exotic from
the $q \bar{q}$ picture. The states that we find qualify as mostly $\rho D^*$
molecules, and fit nicely with the experimental states $D_2^*(2460)$ and 
 $D^*(2640)$. The present study would, thus, suggest for these states a 
different nature than the one usually assumed in terms of $q \bar{q}$ states
\cite{pdg,isgur,close}.


\section{Formalism for $VV$ interaction}
\subsection{Lagrangian}

\begin{figure}
\begin{center}
\includegraphics[width=16cm]{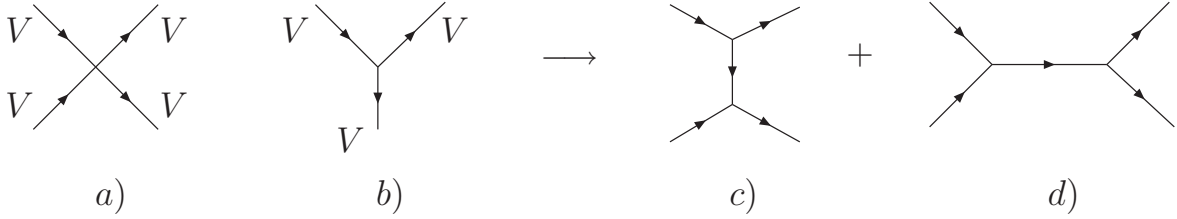}
\end{center}
\caption{Terms of the ${\cal L}_{III}$ Lagrangian: a) four vector contact term,
 Eq.~(\ref{lcont}); b) three-vector interaction, Eq.~(\ref{l3V}); c) $t$ and 
 $u$ channels from vector exchange; d) $s$ channel for vector exchange.}
\label{fig:fig1} 
\end{figure}

We follow the formalism of the hidden gauge symmetry (HGS) for vector mesons of 
\cite{hidden1,hidden2}(see also \cite{hidekoroca} for a practical set of Feynman rules). 
The Lagrangian involving the interaction of 
vector mesons amongst themselves is given by
\begin{equation}
{\cal L}_{III}=-\frac{1}{4}\langle V_{\mu \nu}V^{\mu\nu}\rangle \ ,
\label{lVV}
\end{equation}
where the symbol $\langle \rangle$ stands for the trace in the $SU(4)$ space 
and $V_{\mu\nu}$ is given by 
\begin{equation}
V_{\mu\nu}=\partial_{\mu} V_\nu -\partial_\nu V_\mu -ig[V_\mu,V_\nu]\ ,
\label{Vmunu}
\end{equation}
with $g$ given by
\begin{equation}
g=\frac{M_V}{2f}\ ,
\label{g}
\end{equation}
and $f=93\,MeV$ the pion decay constant. The value of $g$ of Eq.~(\ref{g}) is 
one of the ways to account for the KSFR relation \cite{KSFR} which
is tied to  
vector meson dominance \cite{sakurai}.
The vector field $V_\mu$ is represented by the $SU(4)$ matrix which is
parameterized by 16 vector mesons including 15-plet and singlet of $SU(4)$,
\begin{equation}
V_\mu=\left(
\begin{array}{cccc}
\frac{\rho^0}{\sqrt{2}}+\frac{\omega}{\sqrt{2}}&\rho^+& K^{*+}&\bar{D}^{*0}\\
\rho^-& -\frac{\rho^0}{\sqrt{2}}+\frac{\omega}{\sqrt{2}}&K^{*0}&D^{*-}\\
K^{*-}& \bar{K}^{*0}&\phi&D^{*-}_s\\
D^{*0}&D^{*+}&D^{*+}_s&J/\psi\\
\end{array}
\right)_\mu \ ,
\label{Vmu}
\end{equation}
where the ideal mixing has been taken for $\omega$, $\phi$ and $J/\psi$.
The interaction of ${\cal L}_{III}$ gives rise to a contact term coming for 
$[V_\mu,V_\nu][V_\mu,V_\nu]$
\begin{equation}
{\cal L}^{(c)}_{III}=\frac{g^2}{2}\langle V_\mu V_\nu V^\mu V^\nu-V_\nu V_\mu
V^\mu V^\nu\rangle\ ,
\label{lcont}
\end{equation}
depicted in Fig.~\ref{fig:fig1} a), and on the other hand it gives rise to a three 
vector vertex
\begin{equation}
{\cal L}^{(3V)}_{III}=ig\langle (\partial_\mu V_\nu -\partial_\nu V_\mu) V^\mu V^\nu\rangle
\label{l3V}\ ,
\end{equation}
depicted in Fig.~\ref{fig:fig1} b). This latter Lagrangian gives rise to a
$VV\to VV$ interaction by means of the exchange of one of the vectors, as 
shown in Fig.~\ref{fig:fig1} c).

The $SU(4)$ structure of the Lagrangian allows us to take into account all the
channels within $SU(4)$ which couple to certain quantum numbers. 
In the present work we shall 
present results for the case of the $\rho D^*$ interaction. The formalism is
the same used in \cite{raquel}. Some approximations were made there which make
the formalism handy and reliable, by neglecting the three-momentum of the vector
mesons with respect to their masses. It is interesting to see that with this
approximation one obtains \cite{hidekoroca} from the hidden gauge approach 
the chiral local
Lagrangians which are used to study the interaction of pseudoscalar mesons
among themselves and the pesudoscalar mesons with vector mesons and with baryons
\cite{Gasser:1984gg,ulf}.

\subsection{Four-vector contact interaction}
\begin{figure}
\begin{center}
\includegraphics{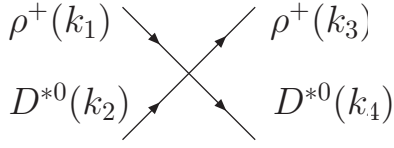}
\end{center}
\caption{Contact term of the $\rho\rho$ interaction.}
\label{fig:fig2}
\end{figure}

Starting with the Lagrangian of Eq.~(\ref{lcont}) we can immediately obtain the 
amplitude of, for instance, $\rho^+ D^{*0}\to \rho^+ D^{*0}$ corresponding to 
Fig.~\ref{fig:fig2}, in the particle base,
\begin{eqnarray}
-it^{(c)}_{\rho^+D^{*0}\to\rho^+D^{*0}}= ig^2 (2\eps^{(1)}_\mu \eps^{(2)}_\nu \eps^{(3)\,\nu} \eps^{(4)\,\mu} -\eps^{(1)}_\mu \eps^{(2)}_\mu \eps^{(3)\,\nu} \eps^{(4)\,\nu} -\eps^{(1)}_\nu \eps^{(2)}_\mu \eps^{(3)\,\nu} \eps^{(4)\,\mu} )\ ,
\label{eq:rhoDvertex}
\end{eqnarray}
where the indices $1,2,3$ and $4$ correspond to the particles with momenta 
$k_1,k_2,k_3$ and $k_4$ in Fig.~\ref{fig:fig2}. 
It is straightforward to write down all amplitudes for other channels.

 In the approximation of neglecting the three-momenta of the vector mesons, only
 the spatial components of the polarization vectors are nonvanishing and then
 one can obtain easily spin projection operators \cite{raquel} into spin
0, 1, 2 states,
  which are given below:
  
  \begin{eqnarray}
{\cal P}^{(0)}&=& \frac{1}{3}\eps_\mu \eps^\mu \eps_\nu \eps^\nu\nonumber\\
{\cal P}^{(1)}&=&\frac{1}{2}(\eps_\mu\eps_\nu\eps^\mu\eps^\nu-\eps_\mu\eps_\nu\eps^\nu\eps^\mu)\nonumber\\
{\cal P}^{(2)}&=&\lbrace\frac{1}{2}(\eps_\mu\eps_\nu\eps^\mu\eps^\nu+\eps_\mu\eps_\nu\eps^\nu\eps^\mu)-\frac{1}{3}\eps_\mu\eps^\mu\epsilon_\nu\epsilon^\nu\rbrace\ ,
\label{eq:projmu}
\end{eqnarray}
  where the order $1,\,2,\,3,\,4$ in the polarization vectors is
understood. We can then write the combination of polarization vectors
appearing in Eq.~(\ref{eq:rhoDvertex}) in terms of the spin combinations
and thus we obtain the kernel of the 
   interaction which will be later on used to solve the Bethe-Salpeter
   equation. However, it is practical to construct the isospin combinations
   before the spin projection is done.

     Recalling that we have an isospin doublet with $(-D^{*0},D^{*+})$ and one isospin triplet, $(\rho^-,\rho^0,-\rho^+)$, the $I=1/2$ and 3/2 combinations are written as 
     
      \begin{eqnarray}
      |\rho D^*,I=1/2,I_3=1/2\rangle&=&\sqrt{\frac{2}{3}}|\rho^+ D^{*0}\rangle-\frac{1}{\sqrt{3}}|\rho^0 D^{*+}\rangle,\nonumber\\
      |\rho D^*,I=1/2,I_3=3/2\rangle&=&\frac{1}{\sqrt{3}}|\rho^+ D^{*0}\rangle+\sqrt{\frac{2}{3}}|\rho^0 D^{*+}\rangle\ .
      \label{eq:isospincomb}
      \end{eqnarray}
We then find the amplitudes in the isospin base by forming linear
combinations of the amplitudes in the particle base weighted by the
Clebsh-Gordan coefficients as given in Eq.~(\ref{eq:isospincomb}),
       \begin{eqnarray}
t^{(I=1/2)}&=&g^2(-\frac{7}{2}\eps_\mu\eps_\nu\eps^\nu\eps^\mu+\frac{5}{2}\eps_\mu\eps^\mu\eps_\nu\eps^\nu+\eps_\mu\eps_\nu\eps^\mu\eps^\nu)\ ,\nonumber\\
t^{(I=3/2)}&=&g^2(\eps_\mu\eps_\nu\eps^\nu\eps^\mu+\eps_\mu\eps_\nu\eps^\mu\eps^\nu-2\eps_\mu\eps^\mu\eps_\nu\eps^\nu)\ .
\label{eq:rhoDIsospin}
\end{eqnarray}
These amplitudes, after
      projection into the spin channels, give rise to the following kernels
      (potential) for $I=1/2$,
      \begin{eqnarray}
      t^{(I=1/2,S=0)}&=&+5 g^2\ ,\nonumber\\
      t^{(I=1/2,S=1)}&=&+\frac{9}{2} g^2\ ,\nonumber\\
      t^{(I=1/2,S=2)}&=&-\frac{5}{2} g^2\ ,
      \label{eq:rhoDIS}
      \end{eqnarray}
and also for the case of $I=3/2$,
\begin{eqnarray}
      t^{(I=3/2,S=0)}&=&-4 g^2\ ,\nonumber\\
      t^{(I=3/2,S=1)}&=&0\ ,\nonumber\\
      t^{(I=3/2,S=2)}&=&+2 g^2\ .
      \label{eq:rhoDIS3}
      \end{eqnarray}

When one or two $\rho$ meson(s) are replaced by the $\omega$ meson(s),
where is only one isospin state $I=1/2$, we find the following
interaction terms,
      \begin{eqnarray}
t^{(I=1/2)}_{\rho D^*\to \omega D^*}&=&\frac{\sqrt{3}}{2}
g^2(2\eps_\mu\eps_\nu\eps^\mu\eps^\nu-\eps_\mu\eps_\nu\eps^\nu\eps^\mu-\eps_\mu\eps^\mu\eps_\nu\eps^\nu)\ ,\nonumber\\
t^{(I=1/2)}_{\omega D^*\to \omega D^*}&=&-\frac{1}{2}g^2(\eps_\mu\eps_\nu\eps^\nu\eps^\mu+\eps_\mu\eps^\mu\eps_\nu\eps^\nu-2\eps_\mu\eps_\nu\eps^\mu\eps^\nu)\ .
\label{eq:rhoDomIsospin}
\end{eqnarray}
After projection in spin they become for $\rho D^*\rightarrow \omega D^*$,
\begin{eqnarray}
      t^{(I=1/2,S=0)}_{\rho D^*\to \omega D^*}&=&-\sqrt{3}g^2\ ,\nonumber\\
      t^{(I=1/2,S=1)}_{\rho D^*\to \omega D^*}&=&+\frac{3\sqrt{3}}{2} g^2\ ,\nonumber\\
      t^{(I=1/2,S=2)}_{\rho D^*\to \omega D^*}&=&+\frac{\sqrt{3}}{2} g^2\ ,
      \label{eq:omrhoDIS}
      \end{eqnarray}
and in the same way, we have for $\omega D^*\to \omega D^*$,
\begin{eqnarray}
      t^{(I=1/2,S=0)}_{\omega D^*\to \omega D^*}&=&- g^2\ ,\nonumber\\
      t^{(I=1/2,S=1)}_{\omega D^*\to \omega D^*}&=&+\frac{3}{2}g^2\ ,\nonumber\\
      t^{(I=1/2,S=2)}_{\omega D^*\to \omega D^*}&=&+\frac{1}{2} g^2\ .
      \label{eq:omDIS}
      \end{eqnarray}
\subsection{$\rho$ meson exchange terms}

\begin{figure}
\begin{center}
\includegraphics{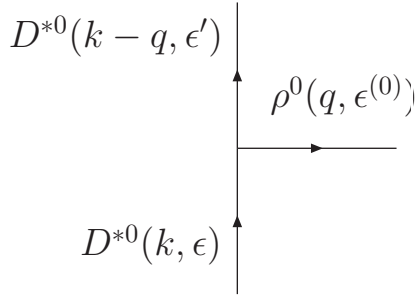}
\end{center}
\caption{three-vector vertex diagram.}
\label{fig:fig5}
\end{figure}

 From the Lagrangian of Eq.~(\ref{l3V}) we get the three-vector vertex as depicted 
in Fig.~\ref{fig:fig5}.
  For practical purposes it is convenient to rewrite the three-vector Lagrangian
  of Eq.~(\ref{l3V}) as,
  \begin{eqnarray}
{\cal L}^{(3V)}_{III}=ig\langle V^\nu\partial_\mu V_\nu V^\mu-\partial_\nu V_\mu
V^\mu V^\nu\rangle \nonumber\\
=ig\langle (V^\mu\partial_\nu V_\mu -\partial_\nu V_\mu
V^\mu) V^\nu\rangle
\label{l3Vsimp}\ .
\end{eqnarray}
In Eq.~(\ref{l3Vsimp}) we have a  three-vector vertex, where any of the three
vector fields can correspond in principle to the exchanged vector in the
diagram of Fig. 1(c). Nevertheless, by assuming that the three-momenta of the
external vectors can be neglected
as compared with
the vector mass,  the
polarization vectors of the external vector mesons have only spatial components.
 Then by looking at the 
Lagrangian of Eq.~(\ref{l3Vsimp})
we see that the field $V^\nu$   cannot
correspond to an external vector meson. Indeed, if this were the case, the $\nu$
index must be spatial and then the partial derivative $\partial_\nu$  
is replaced by
a three-momentum of the vector mesons which is neglected in the approach. Then 
 $V^\nu$ corresponds to the exchanged vector and this simplifies the
 calculation.
The approximation of neglecting the three-momenta of the external vectors
corresponds to the consideration of only the $s$-wave.

\begin{figure}
\begin{center}
\includegraphics{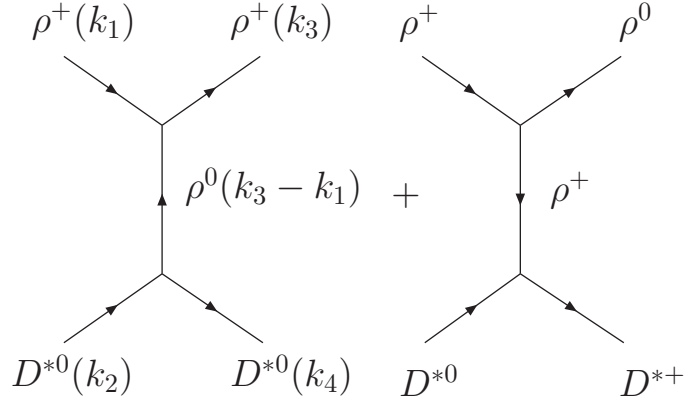}
\end{center}
\caption{Vector exchange diagrams for $\rho D^{*}\to\rho D^{*}$.}
\label{fig:fig6}
\end{figure}

The vertex function corresponding to the diagram of Fig.~\ref{fig:fig5} is given 
by
\begin{eqnarray}
-it^{(3)}=&-&\frac{g}{\sqrt{2}}\lbrace (iq_\mu \eps_\nu^{(0)} -iq_\nu \eps^{(0)}_\mu)\eps'^\mu\eps^{\nu}\nonumber\\&+&(i(k-q)_\mu\eps'_\nu-i(k-q)_\nu\eps'_\mu)\eps^{\mu}\eps^{(0)\nu}\nonumber\\&+&(-ik_\mu \eps_\nu+ik_\nu \eps_\mu)\eps^{(0)\mu}\eps'^\nu\rbrace\ .
\label{vertexfig3}
\end{eqnarray}
With this basic structure we can readily evaluate the amplitude of the first diagram 
of Fig.~\ref{fig:fig6} to obtain 

\begin{eqnarray}
-it^{(ex)}_{\rho^+ D^{*0}\to \rho^+ D^{*0}}=&-&\sqrt{2} g\lbrace (-i(k_3-k_1)_\mu \eps^{(0)}_\nu +i(k_3-k_1)_\nu\eps^{(0)}_\mu)\eps^{(1)\mu}\eps^{(3)\nu}\nonumber\\&+&(-ik_{1\mu}\eps^{(1)}_\nu +ik_{1\nu}\eps^{(1)}_\mu)\eps^{(3)\mu}\eps^{(0)\nu}+(ik_{3\mu}\eps^{(3)}_\nu -ik_{3\nu}\eps^{(3)}_\mu)\eps^{(0)\mu}\eps^{(1)\nu}\rbrace\nonumber\\ &\times&\frac{i}{(k_3-k_1)^2-M_\rho^2+i\eps}\nonumber\\
&\times&-\frac{g}{\sqrt{2}}\lbrace (i(k_2-k_4)_\mu \eps_\nu^{(0)} -i(k_2-k_4)_\nu \eps^{(0)}_\mu)\eps^{(4)\mu}\eps^{(2)\nu}\nonumber\\&+&(ik_{4\mu}\eps^{(4)}_\nu-ik_{4\nu}\eps^{(4)}_\mu)\eps^{(2)\mu}\eps^{(0)\nu}\nonumber\\&+&(-ik_{2\mu} \eps^{(2)}_\nu+ik_{2\nu} \eps^{(2)}_\mu)\eps^{(0)\mu}\eps^{(4)\nu}\rbrace\ .
\label{exchange}
\end{eqnarray}
Recalling that the three-momenta of the external particles is small and neglected, we arrive at the following expression:
\begin{equation}
t^{(ex)}_{\rho^+ D^{*0}\to \rho^+ D^{*0}}=-\frac{g^2}{M_\rho^2}\,(k_1+k_3)\cdot (k_2 +k_4)\,\eps_\mu\eps_\nu\eps^\mu\eps^\nu\ .
\label{eq:exch}
\end{equation}
 By looking at the structure of the second
 diagram
in Fig.~\ref{fig:fig6} we find the
 following result  for the amplitude:
 \begin{equation}
t^{(ex)}_{\rho^+ D^{*0}\to \rho^0 D^{*+}}=\sqrt{2}\,\frac{g^2}{M_\rho^2}\,(k_1+k_3)\cdot (k_2 +k_4)\,\eps_\mu\eps_\nu\eps^\mu\eps^\nu.
\label{eq:exch1}
\end{equation}
We note that the amplitude $\rho^0D^*\rightarrow\rho^0D^*$ 
with $\rho^0$ exchange
vanishes
because the three-$\rho^0$ vertex does not exist due to isospin invariance.
 We can see that in all the cases the combination of vector polarizations is
 the same.
 The isospin projections give us
 \begin{eqnarray}
 t^{(ex,I=1/2)}_{\rho D^{*}\to \rho D^{*}}&=&-2\,\frac{g^2}{M_\rho^2}\,(k_1+k_3)\cdot (k_2 +k_4)\,\eps_\mu\eps_\nu\eps^\mu\eps^\nu\ ,\nonumber\\
 t^{(ex,I=3/2)}_{\rho D^{*}\to \rho D^{*}}&=&\frac{g^2}{M_\rho^2}\,(k_1+k_3)\cdot (k_2 +k_4)\,\eps_\mu\eps_\nu\eps^\mu\eps^\nu\ .
 \label{eq:exchisospin}
 \end{eqnarray}
Now using the equations for the spin projections we can split the terms into
 their spin parts and we obtain
 \begin{eqnarray}
 t^{(ex,I=1/2,S=0,1,2)}_{\rho D^{*}\to \rho D^{*}}&=&-2\,\frac{g^2}{M_\rho^2}\,(k_1+k_3)\cdot (k_2 +k_4)\ ,\nonumber\\
 t^{(ex,I=3/2,S=0,1,2)}_{\rho D^{*}\to \rho D^{*}}&=&\frac{g^2}{M_\rho^2}\,(k_1+k_3)\cdot (k_2 +k_4)\ ,
 \label{eq:exchisospinspin}
 \end{eqnarray}
 that is: we find spin degeneration in the amplitudes which involve the
exchange of one $\rho$ meson. These structures can be simplified using
momentum conservation and one finds: 
 \begin{eqnarray}
 t^{(ex,I=1/2,S=0,1,2)}_{\rho D^{*}\to \rho D^{*}}=&-&2\,\frac{g^2}{M_\rho^2}\,\lbrace\frac{3}{2}s -m^2_\rho-m^2_{D^*}-\frac{(m^2_\rho-m^2_{D^*})^2}{2 s}\rbrace\ ,\nonumber\\
 t^{(ex,I=3/2,S=0,1,2)}_{\rho D^{*}\to \rho D^{*}}=&+&\frac{g^2}{M_\rho^2}\,\lbrace\frac{3}{2}s -m^2_\rho-m^2_{D^*}-\frac{(m^2_\rho-m^2_{D^*})^2}{2 s}\rbrace\ .
 \label{eq:exchisspinsimp}
 \end{eqnarray}
  Note that for the exchange of a vector we do not have contribution from the
 $\omega D^*$ channel. Indeed, the vertex $\omega \omega \omega $ and 
 $\rho \rho \omega $ violate G-parity, and the  the $\rho \omega \omega $
 violates isospin. In the next section we consider the 
 amplitudes which include also the exchange of one heavy vector meson, but we anticipate
 that the amplitudes calculated so far are more relevant than the other ones.
\subsection{$D^*$-exchange terms}
 
\begin{figure}
\begin{center}
\includegraphics{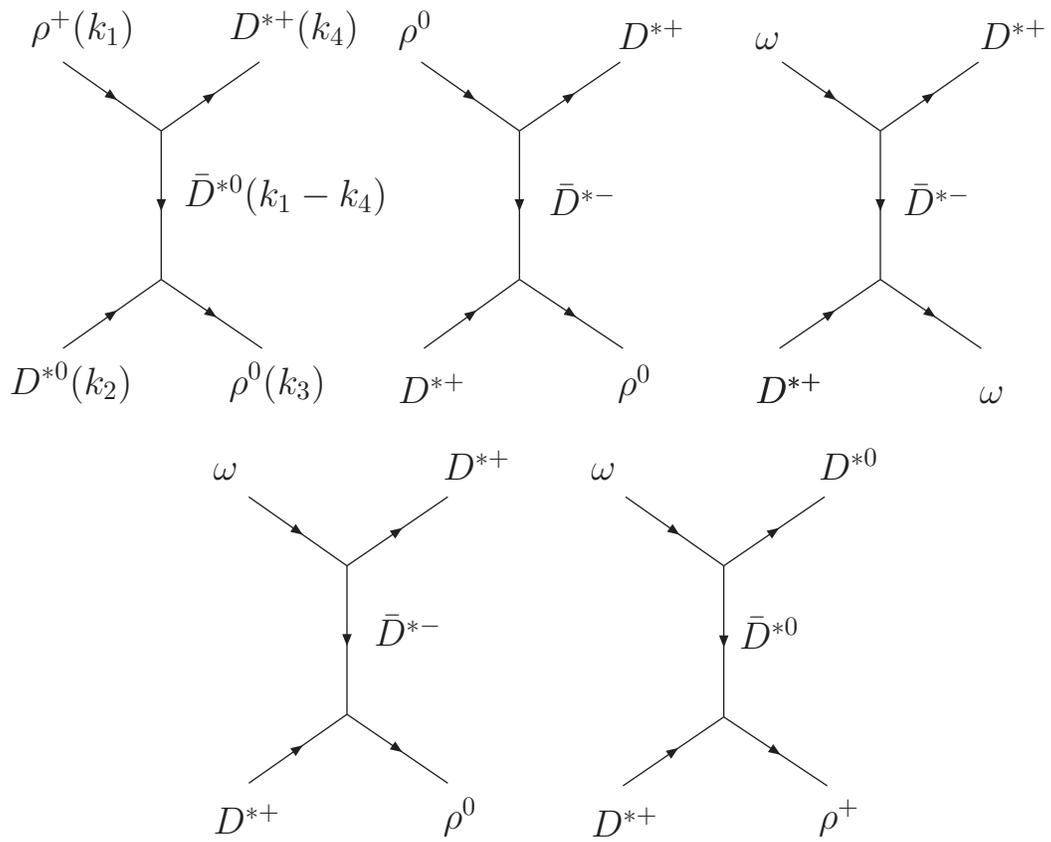}
\end{center}
\caption{Diagrams including the exchange of one heavy vector meson.}
\label{fig:exchD}
\end{figure}

 In this section we are going to take into account the exchange of one
heavy vector meson, $D^*$ or $\bar{D}^*$, in the channels $\rho D^*$
and $\omega D^*$, by means of the diagrams that we draw in
Fig.~\ref{fig:exchD}. 
Note that in Fig.~\ref{fig:exchD} the vertex $\omega\omega D^*$ does not appear because 
it violates isospin.
By means of the Lagrangian of Eq.~(\ref{l3V}) we arrive at the following
amplitudes for the first and second diagrams depicted in that figure, 
\begin{eqnarray}
t^{(D^*-ex)}_{\rho^+ D^{*0}\to \rho^0 D^{*+}}&=&\frac{1}{\sqrt{2}}\frac{g^2}{M_{D^*}^2}\,(k_1+k_4)\cdot (k_2 +k_3)\,\eps_\mu\eps_\nu\eps^\nu\eps^\mu\nonumber\\
t^{(D^*-ex)}_{\rho^0 D^{*+}\to \rho^0 D^{*+}}&=&\frac{1}{2}\frac{g^2}{M_{D^*}^2}\,(k_1+k_4)\cdot (k_2 +k_3)\,\eps_\mu\eps_\nu\eps^\nu\eps^\mu\ .
\label{eq:exchDu}
\end{eqnarray}
The isospin decomposition can be done as before by making weighted sums
of Eq.~(\ref{eq:exchDu}). We find,
\begin{eqnarray}
t^{(D^*-ex,I=1/2)}_{\rho D^{*}\to \rho D^{*}}&=&-\frac{1}{2}\frac{\kappa\, g^2}{M_{\rho}^2}\,(k_1+k_4)\cdot (k_2 +k_3)\,\eps_\mu\eps_\nu\eps^\nu\eps^\mu\nonumber\\
t^{(D^*-ex,I=3/2)}_{\rho D^{*}\to \rho D^{*}}&=&\frac{\kappa\, g^2}{M_{\rho}^2}\,(k_1+k_4)\cdot (k_2 +k_3)\,\eps_\mu\eps_\nu\eps^\nu\eps^\mu.
\label{eq:exchtu}
\end{eqnarray}
where $\kappa=M_\rho^2/M_{D^{*}}^2$. Upon spin projection we find,
\begin{eqnarray}
t^{(D^*-ex,I=1/2,S=0,2)}_{\rho D^{*}\to \rho D^{*}}&=&-\frac{1}{2}\frac{\kappa\, g^2}{M_{\rho}^2}\,(k_1+k_4)\cdot (k_2 +k_3)\nonumber\\
t^{(D^*-ex,I=1/2,S=1)}_{\rho D^{*}\to \rho D^{*}}&=&\frac{1}{2}\frac{\kappa\, g^2}{M_{\rho}^2}\,(k_1+k_4)\cdot (k_2 +k_3)\ ,
\label{eq:exchtuspin12}
\end{eqnarray}
and similarly for $I=3/2$,
\begin{eqnarray}
t^{(D^*-ex,I=3/2,S=0,2)}_{\rho D^{*}\to \rho D^{*}}&=&\frac{\kappa\, g^2}{M_{\rho}^2}\,(k_1+k_4)\cdot (k_2 +k_3)\nonumber\\
t^{(D^*-ex,I=3/2,S=1)}_{\rho D^{*}\to \rho D^{*}}&=&-\frac{\kappa\, g^2}{M_{\rho}^2}\,(k_1+k_4)\cdot (k_2 +k_3)\ .
\label{eq:exchtuspin32}
\end{eqnarray}
Again by neglecting
the three-momenta of the external particles and by taking
the center-of-mass reference system, one can express the factors which
involve momenta as 
\begin{eqnarray}
(k_1+k_3)\cdot (k_2 +k_4)&=&\frac{3}{2}s -m^2_\rho-m^2_{D^*}-\frac{(m^2_\rho-m^2_{D^*})^2}{2 s}\ ,\nonumber\\
(k_1+k_4)\cdot (k_2 +k_3)&=&\frac{3}{2}s -m^2_\rho-m^2_{D^*}+\frac{(m^2_\rho-m^2_{D^*})^2}{2 s}\ .
\label{eq:momenta}
\end{eqnarray}

As we can observe, the spin degeneracy seen in the $\rho$-meson exchange
amplitudes is lost in the $D^*$-exchange amplitudes. However, it is
broken only a little due to the suppressing factor
$\kappa=m_\rho^2/m_{D^*}^2\sim 0.15$. 

The results are summarized in the columns titled ``$D^*$-exchange'' in
Tables~\ref{tab:amplitudes}, 
\ref{tab:amplitudesrhoom} and \ref{tab:amplitudesomom}. These new terms represent corrections of the order of $10\%$ of the $\rho$-exchange ones. 
As can be observed in the total amplitudes, we find attraction in the sector
$I=1/2$, whereas the sector $I=3/2$ turns
out repulsive. It is interesting to see that the exotic $I=3/2$ channel
has a repulsive interaction. This seems to be a rather universal in this
kind of studies \cite{geng,Hyodo:2006kg,Hyodo:2006yk}. 
\begin{table}[h]
 \begin{center}
\begin{tabular}{c|c|c|c|c|c}
$I$&$S$&Contact&$\rho$-exchange&$D^*$-exchange&$\sim$ Total$[I(J^{P})]$\\
\hline
\hline
$1/2$&$0$&$+5g^2$&$-2\frac{g^2}{M_{\rho}^2}\,(k_1+k_3)\cdot (k_2 +k_4)$&$-\frac{1}{2}\frac{\kappa\, g^2}{M_{\rho}^2}\,(k_1+k_4)\cdot (k_2 +k_3) $&$-16g^2[1/2(0^{+})]$\\
\hline
$1/2$&$1$&$+\frac{9}{2}g^2$&$-2\frac{g^2}{M_{\rho}^2}\,(k_1+k_3)\cdot (k_2 +k_4)$&$+\frac{1}{2}\frac{\kappa\, g^2}{M_{\rho}^2}\,(k_1+k_4)\cdot (k_2 +k_3) $&$-14.5g^2[1/2(1^{+})]$\\
\hline
$1/2$&$2$&$-\frac{5}{2}g^2$&$-2\frac{g^2}{M_{\rho}^2}\,(k_1+k_3)\cdot (k_2 +k_4)$&$-\frac{1}{2}\frac{\kappa\, g^2}{M_{\rho}^2}\,(k_1+k_4)\cdot (k_2 +k_3) $&$-23.5g^2[1/2(2^{+})]$\\
\hline
$3/2$&$0$&$-4g^2$&$+\frac{g^2}{M_{\rho}^2}\,(k_1+k_3)\cdot (k_2 +k_4)$&$+\frac{\kappa\, g^2}{M_{\rho}^2}\,(k_1+k_4)\cdot (k_2 +k_3) $&$+8g^2[3/2(0^{+})]$\\
\hline
$3/2$&$1$&$0$&$+\frac{g^2}{M_{\rho}^2}\,(k_1+k_3)\cdot (k_2 +k_4)$&$-\frac{\kappa\, g^2}{M_{\rho}^2}\,(k_1+k_4)\cdot (k_2 +k_3) $&$+8g^2[3/2(1^{+})]$\\
\hline
$3/2$&$2$&$+2g^2$&$+\frac{g^2}{M_{\rho}^2}\,(k_1+k_3)\cdot (k_2 +k_4)$&$+\frac{\kappa\, g^2}{M_{\rho}^2}\,(k_1+k_4)\cdot (k_2 +k_3) $&$+14g^2[3/2(2^{+})]$\\
\hline
\end{tabular} 
\end{center}
\caption{$V(\rho D^*\to \rho D^*)$ for the different spin-isospin channels including the exchange of one 
heavy vector meson. The approximate Total is obtained at the threshold of $\rho D^*$.}
\label{tab:amplitudes}
\end{table} 

\begin{table}[h]
 \begin{center}
\begin{tabular}{c|c|c|c|c|c}
$I$&$S$&Contact&$\rho$-exchange& $D^*$-exchange&$\sim $ Total$[I(J^{P})]$\\
\hline
\hline
$1/2$&$0$&$-\sqrt{3}g^2$&-&$+\frac{\sqrt{3}}{2}\frac{\kappa\, g^2}{M_{\rho}^2}\,(k_1+k_4)\cdot (k_2 +k_3) $&$0[1/2(0^+)]$\\
\hline
$1/2$&$1$&$+\frac{3\sqrt{3}}{2}g^2$&-&$-\frac{\sqrt{3}}{2}\frac{\kappa\, g^2}{M_{\rho}^2}\,(k_1+k_4)\cdot (k_2 +k_3) $&$\frac{\sqrt{3}}{2}g^2[1/2(1^+)]$\\
\hline
$1/2$&$2$&$+\frac{\sqrt{3}}{2}g^2$&-&$+\frac{\sqrt{3}}{2}\frac{\kappa\, g^2}{M_{\rho}^2}\,(k_1+k_4)\cdot (k_2 +k_3) $&$\frac{3\sqrt{3}}{2}g^2[1/2(2^+)]$\\
\hline
\end{tabular} 
\end{center}
\caption{$V(\rho D^*\to \omega D^*)$ for the different spin-isospin channels including the exchange of one 
heavy vector meson. The approximate Total is obtained at the threshold of $\rho D^*$.}
\label{tab:amplitudesrhoom}
\end{table} 

\begin{table}[h]
 \begin{center}
\begin{tabular}{c|c|c|c|c|c}
$I$&$S$&Contact&$\rho$-exchange&$D^*$-exchange&$\sim $ Total$[I(J^{P})]$\\
\hline
\hline
$1/2$&$0$&$-g^2$&-&$+\frac{1}{2}\frac{\kappa\, g^2}{M_{\rho}^2}\,(k_1+k_4)\cdot (k_2 +k_3) $&$0[1/2(0^+)]$\\
\hline
$1/2$&$1$&$+\frac{3}{2}g^2$&-&$-\frac{1}{2}\frac{\kappa\, g^2}{M_{\rho}^2}\,(k_1+k_4)\cdot (k_2 +k_3) $&$\frac{1}{2}g^2[1/2(1^+)]$\\
\hline
$1/2$&$2$&$+\frac{1}{2}g^2$&-&$+\frac{1}{2}\frac{\kappa\, g^2}{M_{\rho}^2}\,(k_1+k_4)\cdot (k_2 +k_3) $&$\frac{3}{2}g^2[1/2(2^+)]$\\
\hline
\end{tabular} 
\end{center}
\caption{$V(\omega D^*\to \omega D^*)$ for the different spin-isospin channels including the exchange of one 
heavy vector meson. The approximate Total is obtained at the threshold of $\rho D^*$.}
\label{tab:amplitudesomom}
\end{table} 
\subsection{T-matrix}
The results obtained in Tables \ref{tab:amplitudes},
\ref{tab:amplitudesrhoom} and \ref{tab:amplitudesomom}  
provide the kernel or potential $V$ to be used in the Bethe-Salpeter
equation 
in its on-shell factorized form,
\begin{equation}
T= \frac{V}{1-VG}\ ,
\label{Bethe}
\end{equation}
for each spin-isospin channel independently. Here $G$ is the two meson loop 
function 
\begin{equation}
G=i\int \frac{d^4 q}{(2\pi)^4}\frac{1}{q^2-m_1^2+i\eps}\frac{1}{(P-q)^2-m_2^2+i\eps}\ ,
\label{loop}
\end{equation}
which upon using dimensional regularization can be recast as
\begin{eqnarray}
G&=&{1 \over 16\pi ^2}\biggr( \alpha +Log{m_1^2 \over \mu ^2}+{m_2^2-m_1^2+s\over 2s}
  Log{m_2^2 \over m_1^2}\nonumber\\ 
  &+&{p\over \sqrt{s}}\Big( Log{s-m_2^2+m_1^2+2p\sqrt{s} \over -s+m_2^2-m_1^2+
  2p\sqrt{s}}+Log{s+m_2^2-m_1^2+2p\sqrt{s} \over -s-m_2^2+m_1^2+  2p\sqrt{s}}\Big)\biggr)\ ,
  \label{dimreg}
\end{eqnarray}
 where $P$ is the total four-momentum of the two mesons, $p$ is the three-momentum 
 of the mesons in the center-of-mass frame. Analogously, using a cut 
 off one obtains
 \begin{equation}
G=\int_0^{q_{max}} \frac{q^2 dq}{(2\pi)^2} \frac{\omega_1+\omega_2}{\omega_1\omega_2 [{(P^0)}^2-(\omega_1+\omega_2)^2+i\epsilon]   } \ ,\label{loopcut}
\end{equation}
 where $q_{max}$ stands for the cut off, $\omega_i=(\vec{q}\,^2_i+m_i^2)^{1/2}$ 
 and the center-of-mass energy ${(P^0)}^2=s$. 
 
 Equation (\ref{Bethe}) in $I=1/2$ is a $2\times 2$ matrix equation with the amplitudes $\rho D\to\rho D$, $\rho D\to \omega D$ and $\omega D\to\omega D$ in the elements $(1,1)$, $(1,2)$  and $(2,2)$. 

\begin{figure}
\begin{center}
\includegraphics{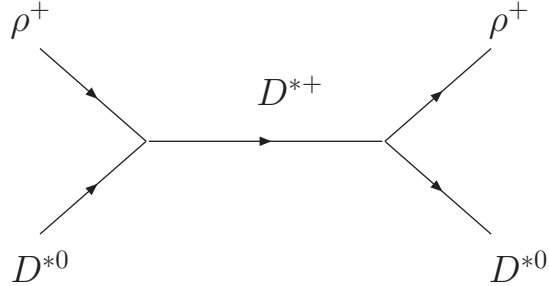}
\end{center}
\caption{$S$-channel $D^*$ exchange diagram.}
\label{fig:fig8}
\end{figure}

The formalism that we are using is also allowed for $s$-channel $\rho$ or $D$ exchange 
and we can have the diagram of Fig.~\ref{fig:fig8}. But we found in
\cite{raquel}
that this leads to a $p$-wave interaction for equal masses of the
vectors, and only to a minor component of $s$-wave in the case of
different masses \cite{geng}. 

\section{Convolution due to the $\rho$ mass distribution}
The strong attraction in the $I=1/2$ and $S=0,1,2$ channels will produce $\rho D^*$ 
bound states with no width within the present treatment so far. However, this is not 
strictly true because the $\rho$ meson has a large width or equivalently a mass 
distribution that allows the states obtained to decay in $\rho D^*$ for the 
low mass components of the $\rho$ mass distribution. To take this into account 
we follow the traditional method which is to convolute the $G$ function for the 
mass distribution of the $\rho$ meson \cite{hidekoroca} replacing the $G$ 
function by $\tilde{G}$ as follows
\begin{eqnarray}
\tilde{G}(s)&=& \frac{1}{N}\int^{(m_\rho+2\Gamma_\rho)^2}_{(m_\rho-2\Gamma_\rho)^2}d\tilde{m}^2_1(-\frac{1}{\pi}) {\cal I}m\frac{1}{\tilde{m}^2_1-m^2_\rho+i\Gamma\tilde{m}_1} G(s,\tilde{m}^2_1,m_{D^*}^2)\ ,
\label{Gconvolution}
\end{eqnarray}
with
\begin{equation}
N=\int^{(m_\rho+2\Gamma_\rho)^2}_{(m_\rho-2\Gamma_\rho)^2}d\tilde{m}^2_1(-\frac{1}{\pi}){\cal I}m\frac{1}{\tilde{m}^2_1-m^2_\rho+i\Gamma\tilde{m}_1}\ ,
\label{Norm}
\end{equation}
where $\Gamma_\rho=146.2\, MeV$ and for $\Gamma\equiv\Gamma(\tilde{m})$ we take the $\rho$ width for the decay into the pions in $p$-wave
\begin{equation}
\Gamma(\tilde{m})=\Gamma_\rho (\frac{\tilde{m}^2-4m^2_\pi}{m^2_\rho-4m^2_\pi})^{3/2}\theta(\tilde{m}-2m_\pi) \ .
\label{gamma}
\end{equation}
The use of $\tilde{G}$ in Eq.~(\ref{Bethe}) provides a width to the states 
obtained as we will see in the next section. 

\section{Results}

When one introduces the amplitudes obtained in Tables
\ref{tab:amplitudes}, \ref{tab:amplitudesrhoom} and
\ref{tab:amplitudesomom} as a kernel $V$ in Eq.~(\ref{Bethe}), one finds
bound states with zero width in the three different cases $I=1/2$ and
$S=0,1,2$. The pole positions are given in Table
\ref{tab:poleposition}. In Eq.~(\ref{dimreg}) we have fixed the value of
$\mu$ as $1500\, MeV$ and we have fine tuned the subtraction constant,
$\alpha$, around its natural value of $-2$ \cite{ollerulf} in order to
get the position of the $S=2$ resonance at its value of the PDG. To
quantify the freedom one has in this fine tuning we quote that the value
of the mass that we obtain using $\alpha=-2$, is $2346\,MeV$. The value
of $\alpha$ for the pole positions given in the Table
\ref{tab:poleposition} is $-1.74$. 
\begin{table}[h]
\begin{center}
\begin{tabular}{c|c|c}
$I$&$S$&$\sqrt{s}~(MeV)$\\
\hline
\hline
$1/2$&$0$&$2592$\\
\hline
$1/2$&$1$&$2611$\\
\hline
$1/2$&$2$&$2450$\\
\hline
\end{tabular}
\end{center}
\caption{Pole positions for the three different cases}
\label{tab:poleposition}
\end{table}

\begin{figure}
\centering
\begin{tabular}{cc}
\includegraphics[width=0.5\textwidth]{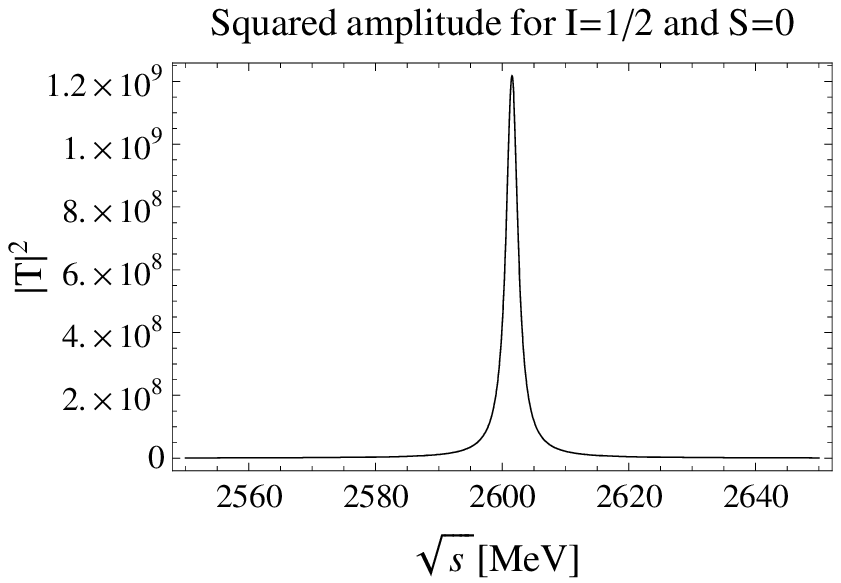}
\includegraphics[width=0.5\textwidth]{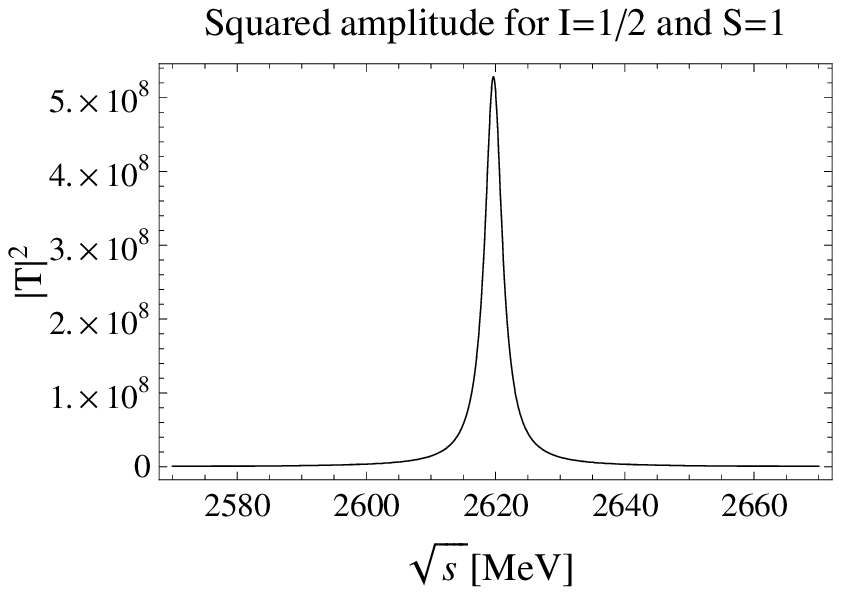}
\end{tabular} 
\caption{Squared amplitude for $I=1/2$ and $S=0,1$ including the convolution of the $\rho$-mass distribution.}
\label{fig:convolucion}
\end{figure}

\begin{figure}
\centering
\begin{tabular}{cc}
\includegraphics[width=0.5\textwidth]{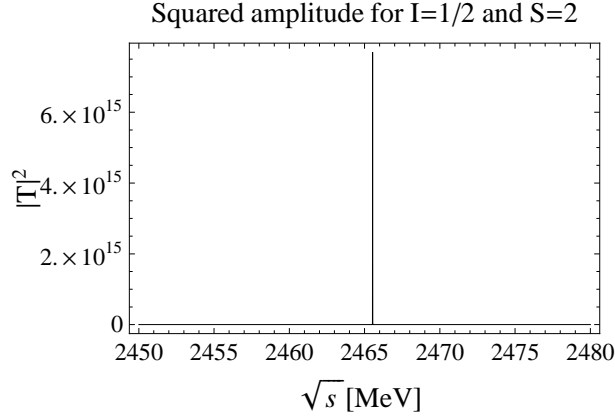}
\end{tabular} 
\caption{Squared amplitude for $I=1/2$ and $S=2$ including the convolution of the $\rho$-mass distribution.}
\label{fig:convolucion1}
\end{figure}

As it was explained in the previous section, the loop function $G$ must
be convoluted to take into account the width of the $\rho$ meson. When
we do it, we find bound states with a small width as we can see in
Figs.~\ref{fig:convolucion} and \ref{fig:convolucion1}. In fact, the
widths found are $\sim$ $5\,MeV$, $4\,MeV$, and $0\,MeV$ for $S=0,1$ and
$2$, respectively. 

In Eq.~(\ref{Bethe}), the amplitude close to a pole looks like 
\begin{equation}
T_{ij}\approx \frac{g_i g_j}{z-z_R}\ ,
\label{poleT}
\end{equation}
where $Re\, z_R$ gives the mass of the resonance and $Im\, z_R$ the half
width. The constants $g_i$ ($i=\rho D^*,\ \omega D^*$), which provide
the coupling of the resonance 
to one particular channel, can be calculated by means of the residues of
the amplitudes as given in Table \ref{res}. 
In the PDG \cite{pdg}, two states are listed, $D^*(2640)$ and
$D_2^*(2460)$ in the sectors $I(J^P)=1/2(?^?)$ and $I(J^P)=1/2(2^+)$,
respectively.
Comparing with the present model predictions as listed in Table
\ref{tab:poleposition},
we
attempt to identify them with those states of $S=1$ and $S=2$, respectively.
The reason to identify the
$D^*(2640)$ with our pole for the case of $S=1$ and not $S=0$ is going
to be clear in the next section when the $D\pi$ channel is
considered. In the case of the $D^*(2640)$ the width quoted in the PDG
is very small, $\Gamma < 15\,MeV$. According to our result after
taking into account the $\rho$ mass distribution, one obtains
3--4 $MeV$, see Fig.~\ref{fig:convolucion}. In the case of 
$D_2^*(2460)$, the width quoted in the PDG is $43\pm 4\,MeV$ for the
$D^{*0}_2$ and $37\pm 6\,MeV$ for $D^{*\pm}_2$. Then, it is clearly not
compatible with the width found here, which is zero, see
Fig.~\ref{fig:convolucion1}, and we need to allow that the resonance
decays to another channel.  For this reason we are going to consider the
$D\pi$ channel in the next section, which is below the threshold of
$\rho D^*$. 
A novelty in this work is that we have obtained a new resonance in the
sector $I=1/2$ and $S=0$ which does not appear in the PDG, see Table
\ref{tab:poleposition}. One should note  that if one takes
Eq.~(\ref{loopcut}) instead of the Eq.~(\ref{dimreg}) with a
cutoff of the order of the natural size $q_{max}=1$--$1.2\,GeV$, the
results are very similar to those of Table \ref{tab:poleposition}
(difference around $1\%$), which is an indication of the stability of the
results. 
\begin{table}[h]
\begin{center}
\begin{tabular}{c|c|c|c}
\hline
Channel&$D_0^*(2600)$ &$D_1^*(2640)$ & $D_2^*(2460)$\\
\hline
\hline
& & &\\
$\rho D^*$&14.32&14.04&17.89\\
\hline
& & &\\
$\omega D^*$&0.53&1.40&2.35\\
\hline
\end{tabular}
\caption{Modules of the couplings $g_i$ in units of GeV for the poles in
 the $S=0,\,1,\,2;\ I=1/2$ sector with the channel $\rho D^*$ and $\omega D^*$.} 
 \label{res}
\end{center}
\end{table}

\section{The $\pi D$ decay mode}
\subsection{Evaluation of the $\pi D$-box diagram}
\begin{figure}
\centering
\includegraphics[width=1\textwidth]{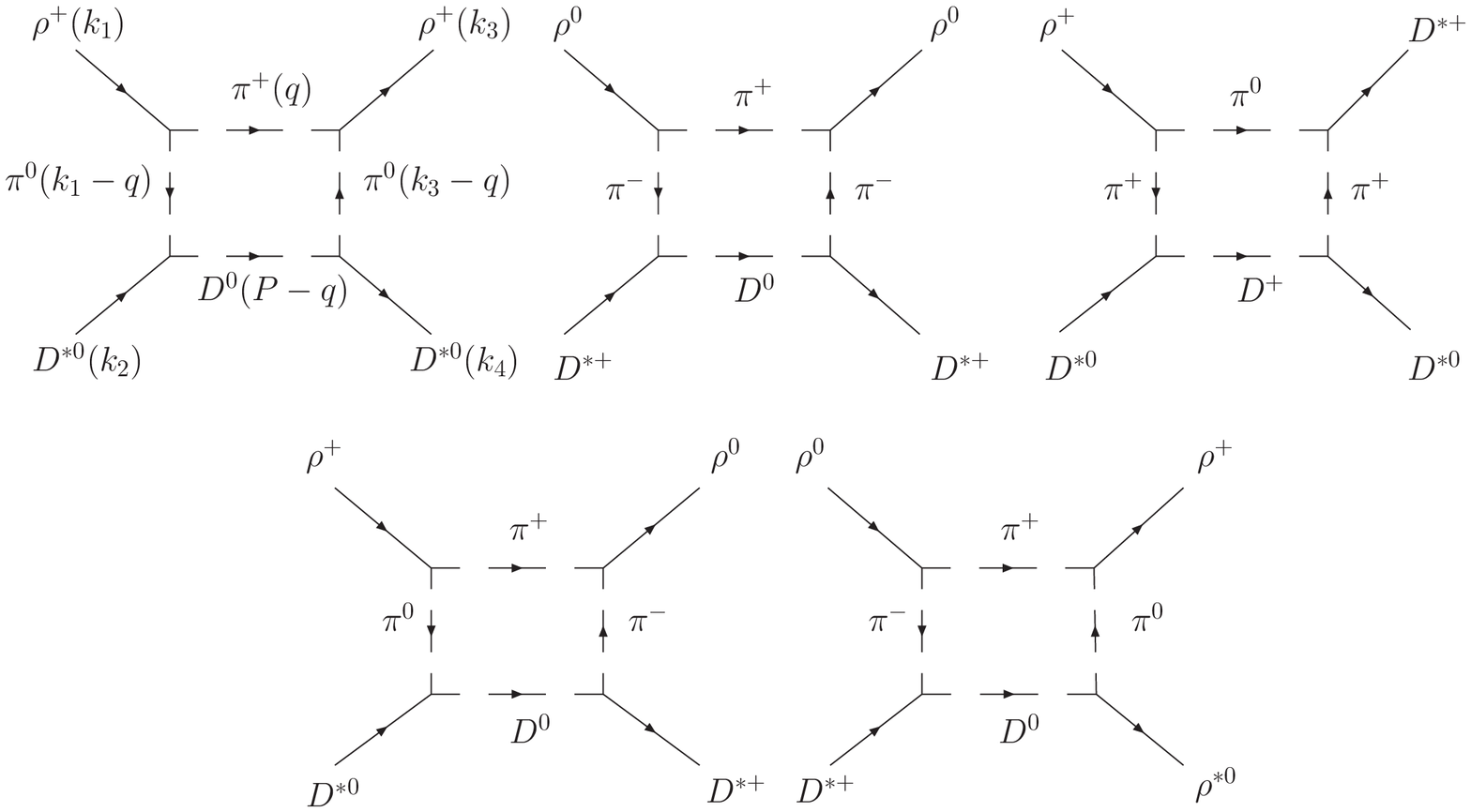}
\caption{$\pi D$-box diagrams}
\label{fig:fig9}
\end{figure}

\begin{figure}
\centering
\begin{tabular}{cc}
\includegraphics[width=0.55\textwidth]{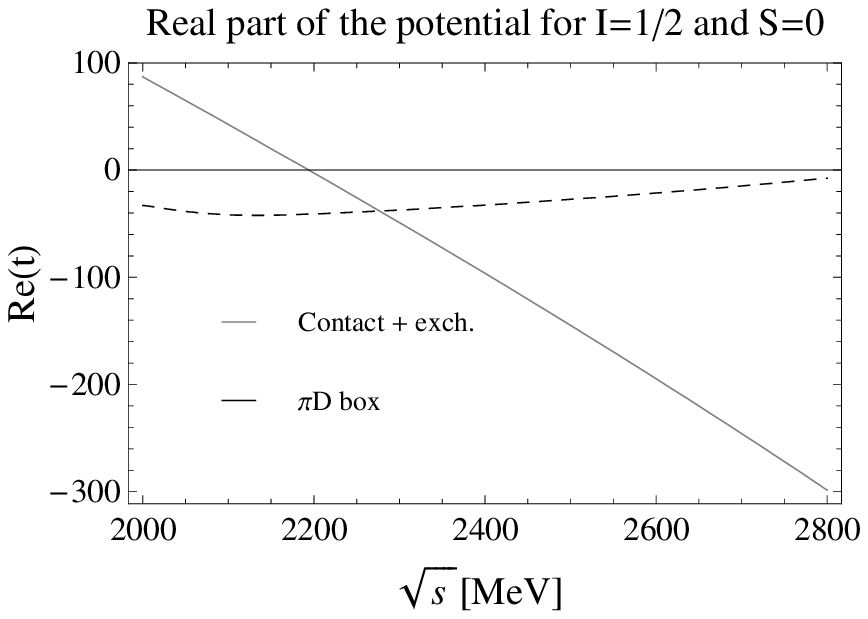}
\includegraphics[width=0.55\textwidth]{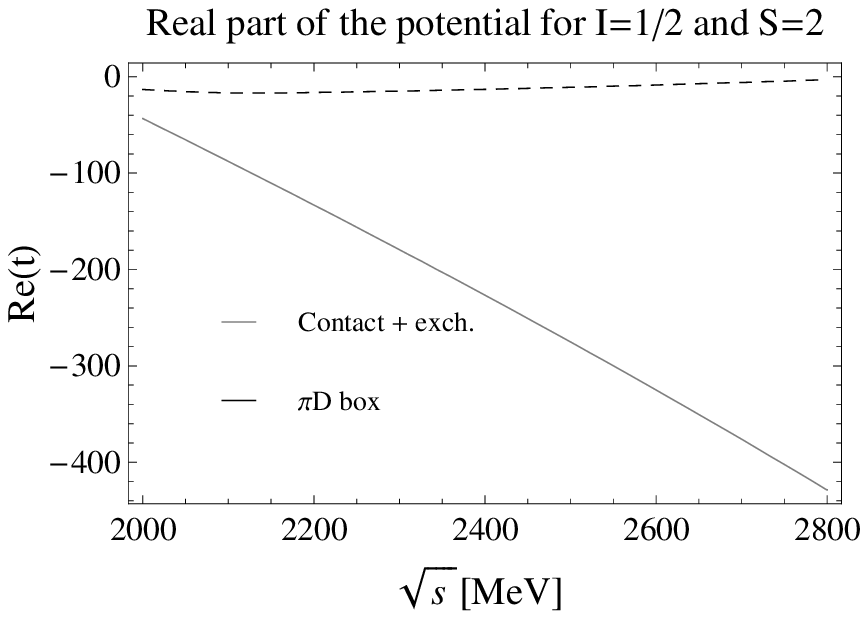}
\end{tabular} 
\caption{Real part of the potential for $I=1/2;\,S=0;$ and $I=1/2;S=2;$.}
\label{fig:boxre}
\end{figure}

\begin{figure}
\centering
\begin{tabular}{cc}
\includegraphics[width=0.55\textwidth]{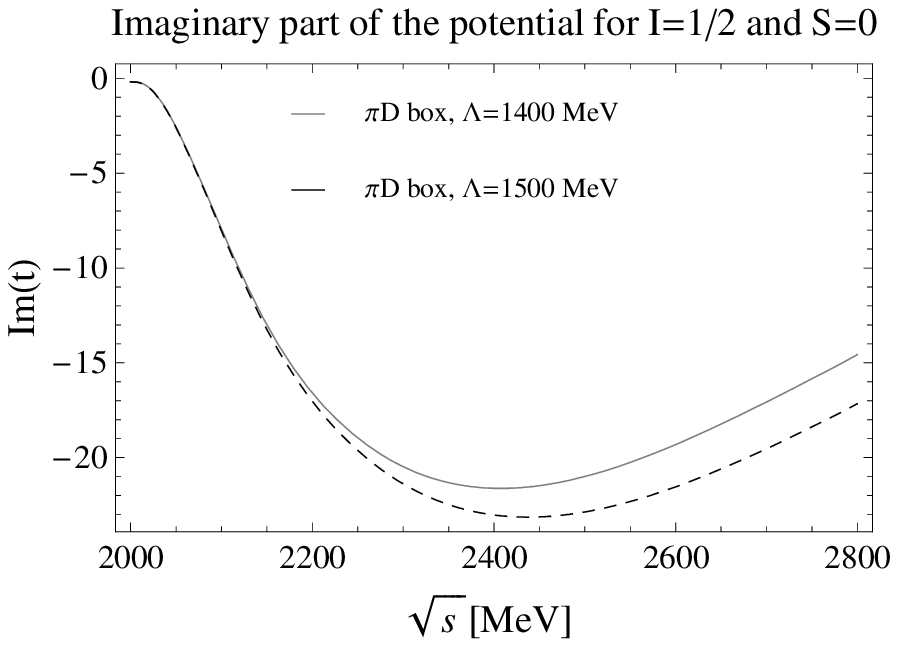}
\includegraphics[width=0.55\textwidth]{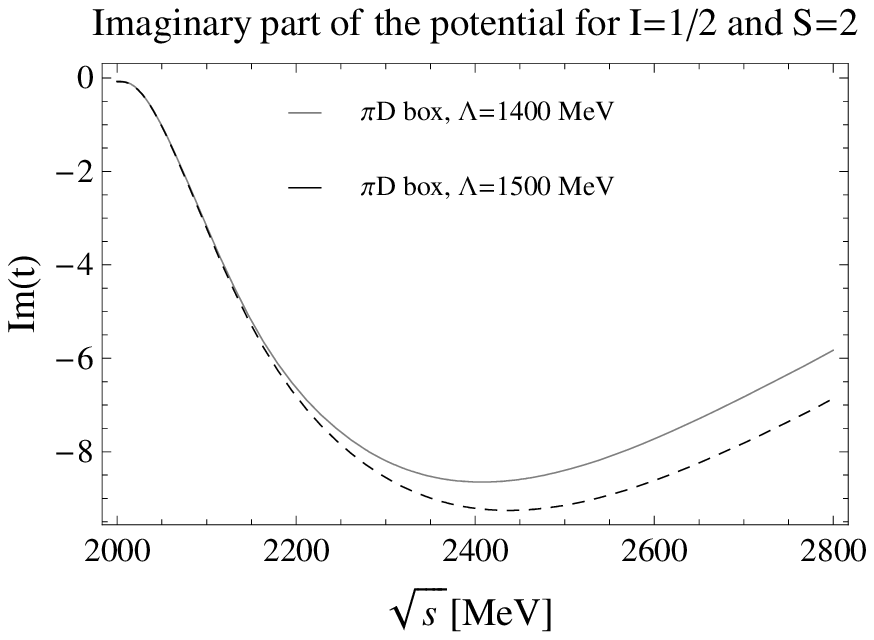}
\end{tabular} 
\caption{Imaginary part of the potential for $I=1/2;\,S=0;$ and $I=1/2;S=2;$.}
\label{fig:boxim}
\end{figure}

In the previous section we have obtained the positions of the poles and
obtained a small width for the states taking into account the finite
width of
$\rho$.
Here we consider a dominant decay mode into the $\pi D$ channel
in order to
give account of a finite width. Our starting point is the set of
diagrams of 
Fig.~\ref{fig:fig9}. One needs the $\rho \pi\pi$ and the $D^* \pi D$
vertex which are provided within the same hidden gauge formalism
\cite{hidden1}, \cite{hidden2}, used in Section $2$, by means of the
Lagrangian 
\begin{equation}
{\cal L}_{V\Phi\Phi}=-ig\langle V^\mu[\Phi,\partial_\mu \Phi]\rangle\ .
\label{lVPP}
\end{equation}
For the first diagram of Fig.~\ref{fig:fig9} we have:
\begin{eqnarray}
-it^{(\pi D)}&=&\int\frac{d^4 q}{(2\pi)^4}(-i)^4\,g^4(\sqrt{2})^2 \,(\frac{1}{\sqrt{2}})^2\,  (k_1-2q)^\mu\eps^{(1)}_\mu \nonumber\\
&\times& (k_3-2q)^\nu\eps^{(3)}_{\nu} (P+k_1-2q)^\alpha\eps^{(2)}_{\alpha}
(P+k_3-2q)^\beta\eps^{(4)}_{\beta}\nonumber\\&\times&\frac{i}{q^2-m^2_\pi+i\eps}\,\frac{i}{(k_1-q)^2-m^2_\pi+i\eps}\nonumber\\
&\times&\,\frac{i}{(P-q)^2-m^2_D+i\eps}\,\frac{i}{(k_3-q)^2-m^2_\pi+i\eps}\ .
\label{tbox}
\end{eqnarray}
Using the approximation that all the polarization vectors are spatial,
it is possible to write the above amplitude as 
\begin{eqnarray}
-it^{(\pi D)}&=&g^4\int\frac{d^4q}{(2\pi)^4}16\,q_i q_jq_lq_m \, \eps^{(1)}_i\eps^{(2)}_j\eps^{(3)}_l\eps^{(4)}_m\nonumber\\&\times&\frac{1}{q^2-m^2_\pi+i\eps}\,\frac{1}{(k_1-q)^2-m^2_\pi+i\eps}\nonumber\\
&\times&\frac{1}{(P-q)^2-m^2_D+i\eps}\,\frac{1}{(k_3-q)^2-m^2_\pi+i\eps}\ .
\label{tbox1}
\end{eqnarray}
This integral is logarithmically divergent and as in \cite{raquel}
we regularize it with a cutoff in the three-momentum of the order of the
natural size,
for which we take
$q_{max}=1.2\,GeV$.
The results do not
change much if one takes a value around this. 
After performing the $dq^0$ integral of Eq.~(\ref{tbox1}),
one finds 
\begin{eqnarray}
V^{(\pi D)}&=&g^4\, \left( \eps^{(1)}_i\eps^{(2)}_i\eps^{(3)}_j\eps^{(4)}_j+\eps^{(1)}_i\eps^{(2)}_j\eps^{(3)}_i\eps^{(4)}_j+\eps^{(1)}_i\eps^{(2)}_j\eps^{(3)}_j\eps^{(4)}_i\right)\nonumber\\
&\times& \frac{8}{15\pi^2}\int^{q_{max}}_0\,dq\, \vec{q}\,^6\,\left(\frac{1}{2\omega}\right)^3\left( \frac{1}{k_1^0+2\omega}\right) ^2\frac{1}{k_2^0-\omega-\omega_D+i\eps}\nonumber\\
&\times&\frac{1}{k_4^0-\omega-\omega_D+i\eps}\,
\frac{1}{k^0_1-2\omega+i\eps}\,\frac{1}{k^0_3-2\omega+i\eps}\,\frac{1}{P^0-\omega-\omega_D+i\eps}\nonumber\\
&\times&\frac{1}{P^0+\omega+\omega_D}\, \left(\frac{1}{k^0_2+\omega+\omega_D}\right)^2\,\frac{1}{2\omega_D}\,f(P^0)
\label{Vbox}
\end{eqnarray}
where
\begin{eqnarray}
f(P^0)=&4&\lbrace-32 k_3^0 P^0 \omega^2 \omega_D ((P^0)^2 - 2 \omega^2 - 
     3 \omega \omega_D - \omega_D^2) \nonumber\\
     &+& 
  2 (k_3^0)^3 P^0 \omega_D ((P^0)^2 - 5 \omega_D^2 - 
     2 \omega \omega_D - \omega_D^2) \nonumber\\&+& 
  (k_3^0)^4 (2 \omega^3 - (P^0)^2 \omega_D + 3 \omega^2 \omega_D + 
     2 \omega \omega_D^2 + \omega_D^3)  \nonumber\\&+&
  4 \omega^2 (8 \omega^5 + 33 \omega^4 \omega_D + 
     54 \omega^3 \omega_D^2 + 3 \omega_D ((P^0)^2 - \omega_D^2)^2 \nonumber\\&+& 
     18 \omega \omega_D^2 (-(P^0)^2 + \omega_D^2) + \omega^2 (-12 \
(P^0)^2 \omega_D + 44 \omega_D^3)) \nonumber\\&-& 
  (k_3^0)^2 (16 \omega^5 + 63 \omega^4 \omega_D + 
     74 \omega^3 \omega_D^2 + \omega_D ((P^0)^2 - \omega_D^2)^2 \nonumber\\&+ &
     32 \omega^2 \omega_D (-(P^0)^2 + \omega_D^2) + \omega (-6 (P^0)^2 \
\omega_D^2 + 6 \omega_D^4))\rbrace
\label{eq:f(p)}
\end{eqnarray}
and $\omega=\sqrt{\vec{q}\,^2+ m_\pi ^2}$, $\omega_D=\sqrt{\vec{q}\,^2+
m_D ^2}$, $P^0=k^0_1+k^0_2$. In Eq.~(\ref{Vbox}) we can see clearly the
sources of the imaginary part in the cuts
$k_2^0(k_4^0)-\omega-\omega_D=0$, $k^0_1(k^0_3)-2\omega=0$,
$P^0-\omega-\omega_D=0$, which give rise to the decays $D^{*0}\to \pi
D$, $\rho \to \pi\pi$ and $\rho D^{*0}\to \pi D$ respectively. As in
\cite{raquel}, to take into account the $\rho$ mass distribution, we
make a simple approach.
%
First we neglect the three-momenta of external particles ($\vec{k}_i\sim
0,\,i=$1, 2, 3, 4) and approximate $k_1^0\sim k_3^0\sim m_\rho$ and 
$k_2^0\sim k_4^0\sim m_{D^*}$. Then we find double poles of
$(1/(k_1^0-2\omega + i\epsilon))^2$ and
$(1/(k_2^0-\omega -\omega_D+ i\epsilon))^2$. These double poles are then
removed by replacing
\begin{equation}
\left(\frac{1}{k_1^0-2\omega+i\epsilon}
\right)^2\rightarrow
\frac{1}{k_1^0-2\omega + i\Gamma_\rho/4}\ 
\frac{1}{k_1^0-2\omega - i\Gamma_\rho/4}
\end{equation}
and so on, considering a finite width for $\rho$ and $D^*$.
Here we set $\Gamma_\rho=146.2\,MeV$ and
$\Gamma_{D^*}=2.1\,MeV$ (results are identical if we put $\Gamma_{D^*}=0\,MeV$). 
Once this is done, the other diagrams of
Fig.~\ref{fig:fig9} are calculated easily, 
which takes into account the decay into the $\pi D$ channel.
One must make a projection into a proper isospin and spin. For isospin,
only $I=1/2$ is allowed, while for spin, $S=0$ and 2 are allowed.
Decay into $S=1$ is forbidden since
the parity of the $\rho D^*$
system for $\rho$ in $s$ wave is positive, which forces the $\pi D$
system to be in $L=0,2$. Since the $\pi$ and $D$ have no spin, the total angular momentum $J$ is equal to $L$ in this case. Therefore, only the $0^+$, $2^+$ quantum
numbers have this decay channel. This provides an explanation on why the
$D^*(2640)$ does not have practically width and the $D^*_2(2460)$ has a
bigger width.  Finally, we obtain 
\begin{eqnarray}
t^{(2\pi,I=1/2,S=0)}&=&20\,\tilde{V}^{(\pi D)}\ ,\nonumber\\
t^{(2\pi,I=1/2,S=2)}&=&8\,\tilde{V}^{(\pi D)}\ ,
\label{eq:boxsis}
\end{eqnarray}
where $\tilde{V}^{(\pi D)}$ is given by Eq.~(\ref{Vbox}) after removing the
polarization vectors. As in \cite{raquel} we use a form factor for a
off-shell $\pi$ in each vertex, which is 
\begin{equation}
F(q)=\frac{\Lambda^2-m^2_\pi}{\Lambda^2+\vec{q}\,^2}\ ,
\label{formfactor}
\end{equation}
with $\Lambda=1400,1500\,MeV$. The real and imaginary parts of the
potential for the contributions that we have calculated
are
plotted in Figs.~\ref{fig:boxre} and \ref{fig:boxim}. As we can see, the
real part of the potential coming from the $\pi D$-box diagram is much
smaller than the real part of the potential coming from the contact plus
exchange terms. 

\subsection{Results with $V^{(\pi D)}$}

\begin{figure}
\centering
\begin{tabular}{cc}
\includegraphics[width=0.52\textwidth]{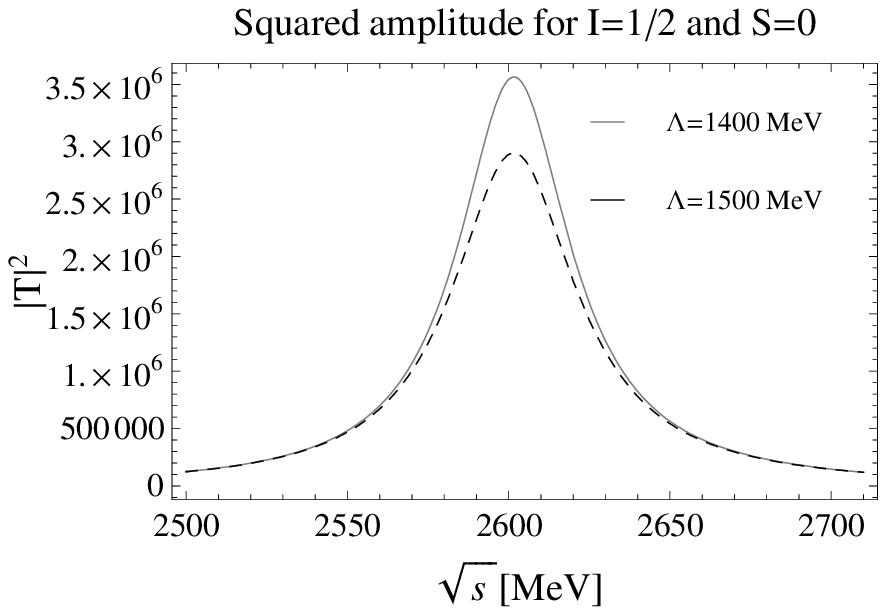}
\includegraphics[width=0.52\textwidth]{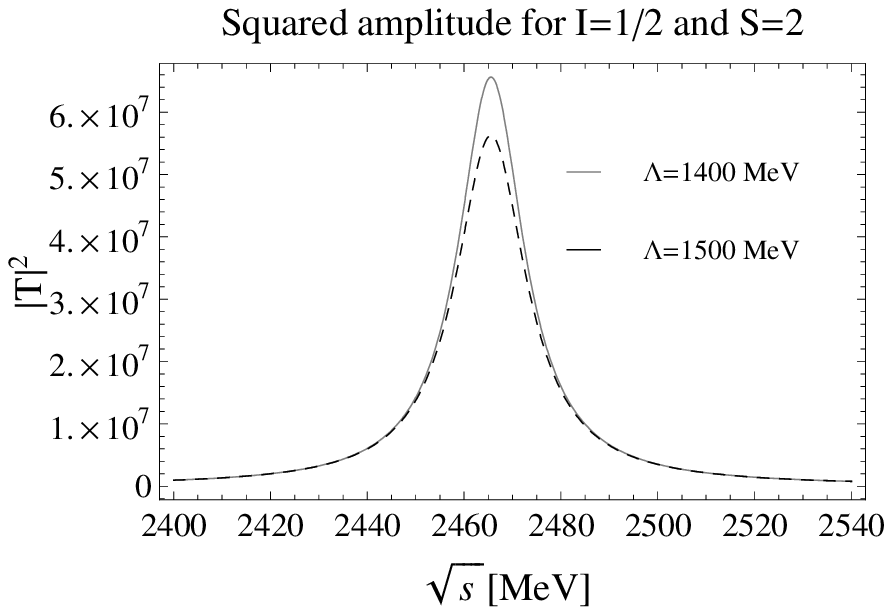}
\end{tabular} 
\caption{Squared amplitude for $S=0$ and $S=2$ including the convolution of the $\rho$-mass distribution and the $\pi D$-box diagram.}
\label{fig:amplit}
\end{figure}

In Fig.~\ref{fig:amplit} we show the results when one introduces the
amplitudes of the Tables \ref{tab:amplitudes},
\ref{tab:amplitudesrhoom}, \ref{tab:amplitudesomom} and the given ones
in Eqs. (\ref{eq:boxsis}) and (\ref{Vbox}) in the Bethe-Salpeter
equation Eq. (\ref{Bethe}). 
 As one can see, now the states for $S=0$ and $S=2$ have larger width
since they can decay to $\pi D$ also. We have found a width of $20\,MeV$
for the $D^*_2(2460)$, which is about $50\%$ of the total width quoted
in the PDG \cite{pdg}. One could also have the $D^*\pi$ decay channel,
which would be possible by means of a anomalous coupling but, as it was
seen in \cite{raquel}, this leads to a smaller contribution than the
other one. Also in the PDG the most important contribution comes from
the $\pi D$ channel. For the case of the new state found the width
obtained is $50\,MeV$ with $\Lambda=1500\,MeV$. In the next section we
introduce new elements of phenomenology that help obtain a somewhat
larger width. 

\subsection{Results with $V^{(\pi D)}$ using a different form factor and the experimental coupling constant $g_{D^*D\pi}$ }

In this section we would like to evaluate again the $\pi D$-box diagram,
but, taking into account 
the strong coupling $g_{D^*D\pi}$ and a different form factor also for an
off-shell pion. 
After the recent measurements by the CLEO Collaboration \cite{Ahmed}, it
is known that the $D^*$ meson couples strongly to $D\pi$.  The
experimental value of this coupling turns out to be larger, almost by a
factor two, than the value obtained from some of the theoretical
predictions using different approaches of the QCD sum rule
\cite{Belyaev,Colangelo}.  
Within the hidden gauge formalism, the vertex $D^{*+}D^0\pi^-$ obtained
from Eq.~(\ref{lVPP}) is 
\begin{equation}
\langle D^{*+}(p)\pi^-(q)\vert  D^0(p+q)\rangle=-2\,g'_{D^*D\pi} q_\mu \eps^\mu\ ,
\label{eq:D*Dpivertex}
\end{equation}
with $g'_{D^*D\pi}=m_{D^*}/2 f_D=6.3$, which is also smaller than
the experimental value for this, $g^{exp}_{D^*D\pi}=8.95\pm 0.15\pm
0.95$. In \cite{Navarra}, the $D^*D\pi$ form factor is evaluated using
the 
QCD sum rule for a $D$ or a $\pi$ off-shell. A parameterization for an
off-shell pion in terms of a form factor
\begin{eqnarray}
F'(q^2)=g_{D^*D\pi} e^{q^2/\Lambda^2}\hspace{0.5cm}\mathrm{with}\hspace{0.2cm}\Lambda=1.2\,GeV\ ,
\label{eq:formfactorprim}
\end{eqnarray}
is taken, together with another form factor to account for off-shell $D$
mesons, which we do not need here since we are concerned about the
imaginary part of the box diagram where the $D$ meson will be on-shell.
In Eq.~(\ref{eq:formfactorprim}) $q^2$ is a four momentum square
$q^2={q^0}^2-\vec{q}^2$ .
With these assumptions on the form factors, a value of
$g_{D^*D\pi}=2\,g'_{D^*D\pi}=14.9$ is obtained in \cite{Navarra}, which
would be in better agreement with experiment ($g'_{D^*D\pi}=7.45$). 

For the first diagram of Fig.~\ref{fig:fig9}, we consider the $q^0$
component of the $\pi^+$ on-shell, hence
$q_0=\frac{s+m^2_\pi-m^2_D}{2\,\sqrt{s}}\sim 769.4\,MeV$ and
$(k^0_1-q^0)\sim 6\,MeV$, for $\rho D^*$ at threshold, in the approximation 
of momentum zero for external particles.
This 
leads to $(k^0-q^0)^2/\Lambda^2\sim 10^{-5}$,
for values of $\Lambda$ around $1\,GeV$. Then it is licit to use the
three-momentum version of the form factor of the
Eq.~(\ref{eq:formfactorprim}) 
for an off-shell pion in each of the vertices, that is, we replace in
Eq.~(\ref{Vbox}) the factor $g^4$ by 
\begin{equation}
g^2_{\rho\pi\pi}\, (g^{exp}_{D^*D\pi})^2 \,(e^{-\vec{q}\,^2/\Lambda^2})^4\ ,
\label{eq:fo1}
\end{equation}
 with $g_{\rho\pi\pi}=m_\rho/2\,f_\pi =4.2$ and
$g^{exp}_{D^*D\pi}=8.95\,MeV$ (the experimental value), $\Lambda\sim
1$--$1.2\,GeV$ and $\vec{q}$ running in the integral. 

\begin{figure}
\centering
\includegraphics[width=1.1\textwidth]{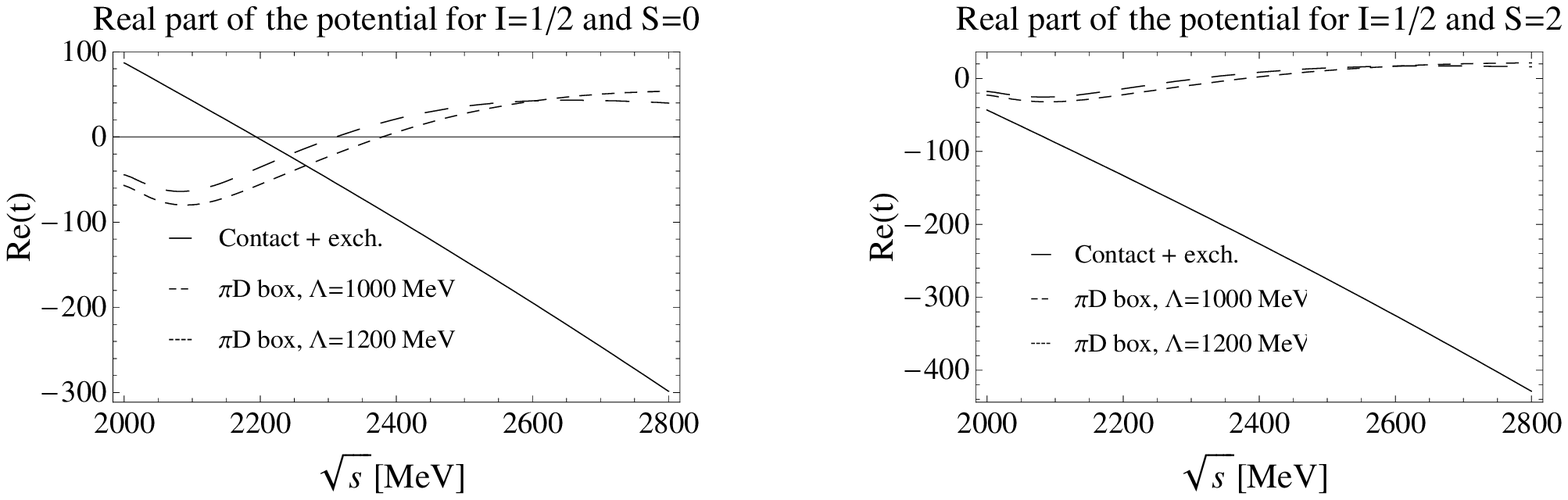}
\caption{Real part of the potential for $I=1/2;\,S=0;$ and $I=1/2;S=2$.}
\label{fig:boxrefo1}
\end{figure}

\begin{figure}
\centering
\includegraphics[width=1.1\textwidth]{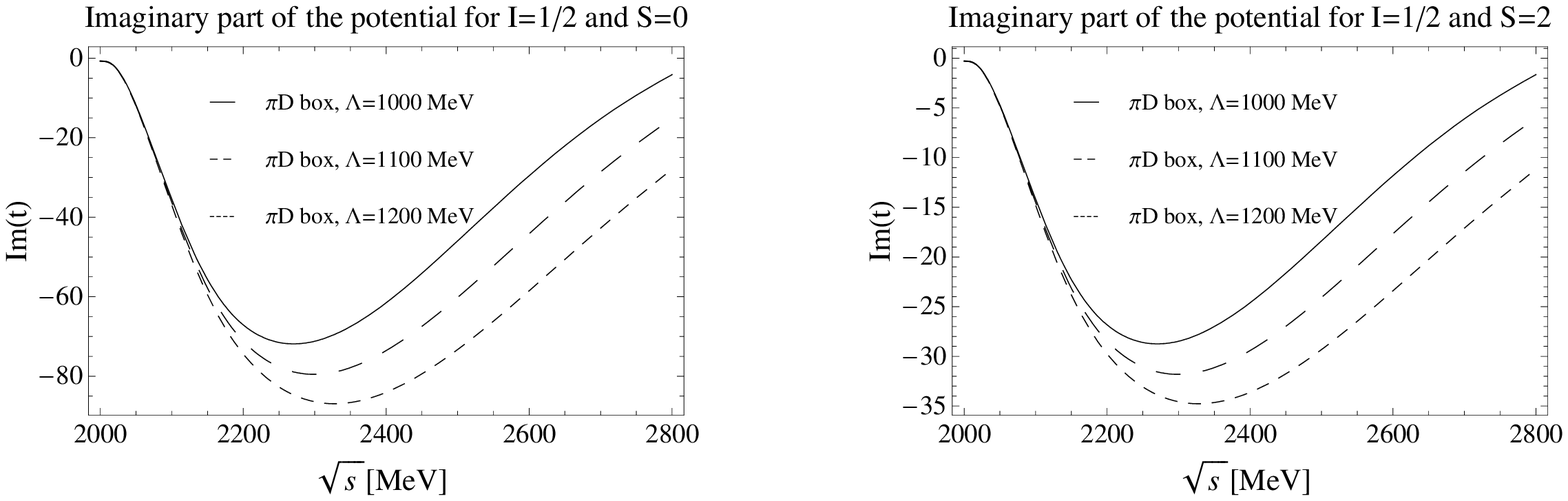}
\caption{Imagiary part of the potential for $I=1/2;\,S=0;$ and $I=1/2;S=2$.}
\label{fig:boximfo1}
\end{figure}

\begin{figure}
\centering
\begin{tabular}{cc}
\includegraphics[width=0.53\textwidth]{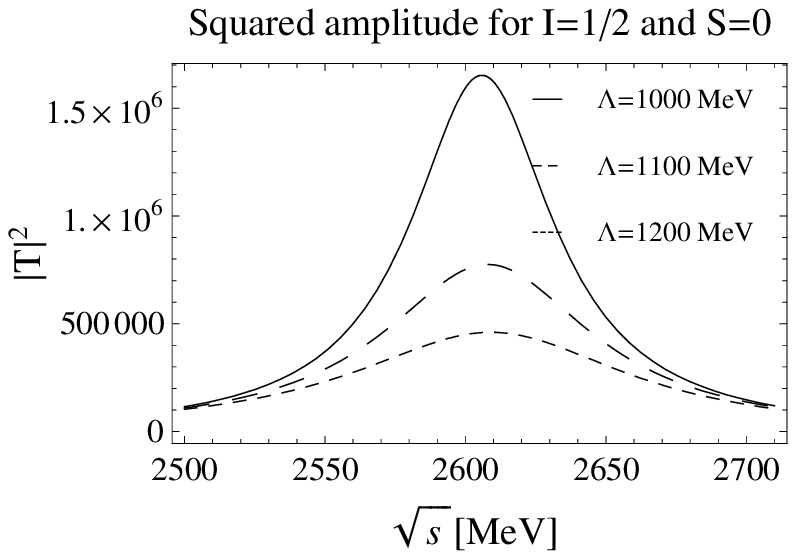}
\includegraphics[width=0.53\textwidth]{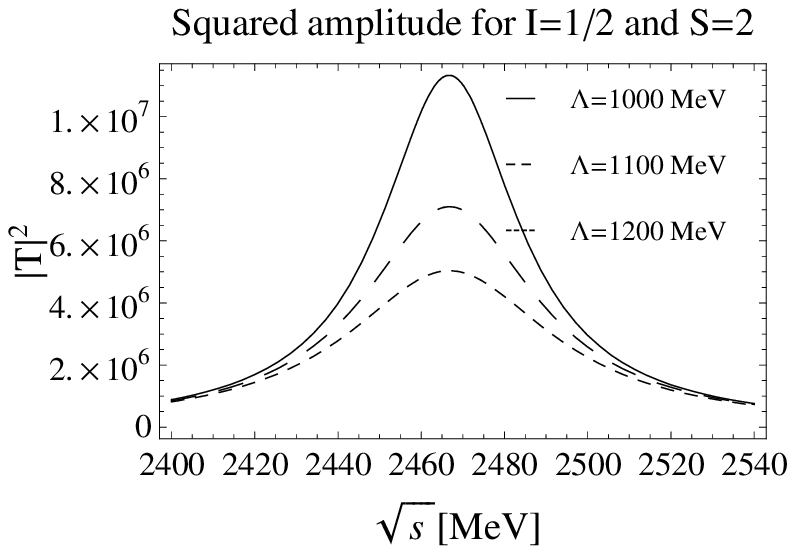}
\end{tabular} 
\caption{Squared amplitude for $S=0$ and $S=2$ including the convolution of the $\rho$-mass distribution and the $\pi D$-boxed diagram.}
\label{fig:tfo1}
\end{figure}

 In Figs.~\ref{fig:boxrefo1} and \ref{fig:boximfo1} we show the real and
imaginary parts of the potential using Eqs.~(\ref{Vbox}) and
(\ref{eq:fo1}). As we can see, the real part of the $\pi D$-box diagram
is very small compared with
those
 coming from the contact term
plus vector exchange terms, and
therefore 
we can ignore it, thus focusing our
attention at the imaginary part. The imaginary part is now larger than
that
in Fig.~\ref{fig:boxim}, but is still comparable with the values quoted
in the PDG \cite{pdg} for the width. We show 
the $\vert T
\vert^2$ in Fig.~\ref{fig:tfo1}
for various cutoff parameters.
The $\vert T\vert^2$ is similar to
Fig.~\ref{fig:amplit}, but now the width is 
slightly
 larger. In the case
of $\Lambda=1\,GeV$, we obtain $40\,MeV$ for $S=2$, very close to the
value quoted by the PDG ($43\pm 4\,MeV$), and $61\,MeV$ for $S=0$. For
$\Lambda=1.2\,GeV$ we obtain for $S=2$ a width around $60\,MeV$.
Therefore, the two form factors, Eq.~(\ref{formfactor})
and Eq.~(\ref{eq:fo1}), provide reasonable values of the width, with a
preference for the option using the experimental $g^{exp}_{D^*D\pi}$
value and $\Lambda=1\,GeV$ in Eq.~(\ref{eq:fo1}). 
 
\section{Conclusions}

We have made a study of the $(\rho\omega) D^*$ interaction using the hidden gauge
formalism. The interaction comes from contact terms plus $\rho$ meson exchange
in the $t$-channel. Of all spin and isospin allowed channels in $s$-wave, we
found strong attraction, enough to bind the system, in $I=1/2,S=0$, $I=1/2,S=1$
and $I=1/2,S=2$.
We also found that in the case of $I=1/2,S=2$ the interaction was more attractive,
than in the other two cases, leading to a tensor state more bound than the scalar
and the axial vector. The consideration of the
$\rho$ mass distribution gives a width to the three states, rather small
in all cases. Consideration of the $\pi D$ decay channel, in an equivalent way
to what was done in the case of the $\rho \rho $ interaction going to $\pi \pi $
in \cite {raquel}, makes the widths larger
and realistic. 
Yet, the smaller phase space
available here makes this contribution relatively smaller than in the case
of the  $\rho \rho $ interaction. We found that the tensor state obtained
matches the properties of the tensor state $D_2^*(2460)$. We predict two more
states with $S=0$ and $S=1$, which are less bound than the tensor state. We find in
the PDG the state $D^*(2640)$ without experimental spin and parity assigned, but
we conjecture that this state should be the $S=1$ state found by us
because we could find a natural explanation for the small experimental
width of this  
state. The other state nearly degenerate in energy with this one, but with spin $S=0$, would
still be to be found. The results obtained here should stimulate the search for
more $D$ states in the region of $2600\,MeV$.

\section*{Acknowledgments}

This work is partly supported by DGICYT contract number
FIS2006-03438. We acknowledge the support of the European Community-Research Infrastructure
Integrating Activity
Study of Strongly Interacting Matter (acronym HadronPhysics2, Grant Agreement
n. 227431)
under the Seventh Framework Programme of EU.
A.~H. is supported in part by the Grant for Scientific Research
Contract No.~19540297 from the  Ministry of Education, Culture,
Science and Technology, Japan.  
H.~N. is supported by the Grant for Scientific Research
No.~18-8661 from JSPS.

\end{document}